\documentclass{ckm}                 

\usepackage{txfonts}            

\confname{Workshop on the CKM Unitarity Triangle, IPPP Durham, April
  2003}

\title{Theoretical Aspects of 
Semileptonic $B$ 
decays\thanks{Talk given at CKM\,-2003, April 5-8 2003, Durham, UK;
Ringberg Phenomenology Workshop, April 27 - May 2, 2003, Ringberg
Castle, Rottach-Egern, Germany; and the $10^{36}$ SLAC Workshop, May
8-10 2003.}}

\author{Nikolai Uraltsev} 
\address{~INFN, Sezione di Milano, Milan, Italy}

\newcommand{\beq}{\begin{equation}}
\newcommand{\eeq}{\end{equation}}
\newcommand{\bea}{\begin{eqnarray}}
\newcommand{\eea}{\end{eqnarray}}

\newcommand{\gsim}{\lower.7ex\hbox{$
\;\stackrel{\textstyle>}{\sim}\;$}}
\newcommand{\lsim}{\lower.7ex\hbox{$
\;\stackrel{\textstyle<}{\sim}\;$}}

\def\lsim{\mathrel{\rlap{\lower3pt\hbox{\hskip0pt$\sim$}}
    \raise1pt\hbox{$<$}}}         
\def\gsim{\mathrel{\rlap{\lower4pt\hbox{\hskip1pt$\sim$}}
    \raise1pt\hbox{$>$}}}         

\renewcommand{\Im}{{\rm Im}\,}

\newcommand{\bibit}[1]{\bibitem{#1}}

\newcommand{\aver}[1]{\langle #1\rangle}

\newcommand{\La}{\overline{\Lambda}}
\newcommand{\Lam}{\Lambda_{\rm QCD}}
\newcommand{\Si}{\overline{\Sigma}}

\newcommand{\GeV}{\,\mbox{GeV}}
\newcommand{\MeV}{\,\mbox{MeV}}
\newcommand{\msp}[1]{\mbox{\hspace*{#1mm}~}}

\newcommand{\matel}[3]{\langle #1|#2|#3\rangle}
\newcommand{\state}[1]{|#1\rangle}

\newcommand{\vep}{\varepsilon}

\begin{document}

\begin{abstract}
Strong interactions are addressed in connection to extracting
$|V_{cb}|$ and determining heavy quark parameters.
A comprehensive approach allows a robust analysis 
not relying on a $1/m_c$ expansion; 
a percent defendable accuracy in $|V_{cb}|$ becomes realistic. 
Some of the heavy quark parameters are already accurately known. 
We have at least one nontrivial precision check of the OPE at the
nonperturbative level in inclusive decays. The alleged controversy
between theory and BaBar data on $\aver{M_X^2}$ is argued to be an
artefact of oversimplifying 
the OPE for high cuts in lepton energy. 
Consequences of the proximity to the `BPS' limit are addressed and
their accuracy qualified. It is suggested that theory-wise 
$B\!\to\!D\,\ell\nu$ near
zero recoil offers an accurate way for measuring $|V_{cb}|$.
\vspace*{-62mm}\\  
\begin{flushright}
{\normalsize Bicocca-FT-03-16\\
UND-HEP-03-BIG\hspace*{.08em}01\\
hep-ph/0306290\vspace*{37.2mm}}\\
\end{flushright}
\end{abstract}

\maketitle

Total semileptonic decay rate $\Gamma_{\rm sl}(B)$ is now
one of the best measured quantities in $B$ physics. This opens an 
accurate way to extract the CKM angle $|V_{cb}|$ where systematic
uncertainties can be thoroughly studied. 
Nonperturbative effects are theoretically controlled by the QCD theorem
\cite{qcdtheor} which established absence of the leading $\Lam/m_b$
power corrections to total decay rates. It is based on OPE and 
applies to all
sufficiently inclusive decay probabilities, semileptonic as well as
nonleptonic or radiatively induced.  
The theorem relates the inclusive $B$ widths
to (short-distance) quark masses and expectation values of local
$b$-quark operators in actual $B$ mesons. To order $1/m_b^2$ they are
kinetic and chromomagnetic averages $\mu_\pi^2$, $\mu_G^2$
having a straightforward quantum-mechanical meaning.
To order $1/m_b^3$ two new operators, Darwin and $LS$ appear for 
$b\!\to\! c\, \ell\nu\,$ with the averages 
$\rho_D^3$ and $\rho_{LS}^3$, also analogous
to those well known from atomic theory.
The general expansion parameter for inclusive decays is energy 
release, in $b\!\to \!c\,\ell\nu$
it constitutes $m_b\!-\!m_c\simeq 3.5\GeV$. For a recent review, see
Ref.~\cite{ioffe}. 

Within the OPE approach calculating everything sufficiently
inclusive has become a technical matter (although not necessary simple
as illustrated by perturbative corrections). Yet care must be
taken to treat properly certain subtleties; some of them will be
discussed. 

Heavy quark masses are full-fledged QCD parameters entering various
hadronic processes. The expectation values like $\mu_\pi^2$
likewise enjoy the status of observables. The masses and
relevant nonperturbative parameters  can be
determined from the $B$ decay distributions 
themselves \cite{motion,optical}. Nowadays this
strategy is being implemented in a number of experimental studies.

Some progress has been recently made on the theoretical side as
well. A resummation of the dominant perturbative
corrections in the context of Wilsonian OPE was performed for the
semileptonic width and showed safe convergence \cite{blmvcb,imprec}, 
eliminating one of the major theoretical uncertainties. Higher-order
nonperturbative corrections were studied and a new class of
nonperturbative effects identified associated with the nonvanishing
expectation values of the four-quark operators with charm field, 
$\matel{B}{\bar{b}c\,\bar{c}b}{B}$ \cite{imprec}. The so-called
``BPS'' regime was proposed as an approximation based on the hierarchy
$\mu_\pi^2\!-\!\mu_G^2 \!\ll\! \mu_\pi^2$ suggested by experiment, and
its immediate consequences analyzed \cite{chrom}. The
new results extending them will be addressed in the end.

The new generation of data provides accurate measurements of
many inclusive characteristics in $B$ decays. At the same time, 
the proper theoretical formalism gradually finds its way into their 
analyses. To summarize the current stage:\\
$\bullet$ We have an accurate and reliable determination of some heavy
quark parameters directly from experiment\\
$\bullet$ Extracting $|V_{cb}|$ from $\Gamma_{\rm sl}(B)$ has good
accuracy and solid grounds\\
$\bullet$ We have at least one nontrivial precision check of the OPE at
the nonperturbative level.\\ 
The latter comes through comparison of the average lepton energy and
the invariant hadronic mass, as explained in the following section.

Present theory allows to aim a percent accuracy in $|V_{cb}|$.
Such a precision becomes possible owing to a number of theoretical
refinements. 
The low-scale running masses $m_b(\mu)$, $m_c(\mu)$, the
expectation values $\mu_\pi^2(\mu)$, $\mu_G^2(\mu)$, ... are completely
defined and can be determined from experiment with an in principle
unlimited accuracy. Violation of local duality potentially
limiting theoretical predictability, has been scrutinized and found 
to be negligibly small in total semileptonic
$B$ widths \cite{vadem}; it can be controlled experimentally with
dedicated high-statistics measurements.
Present-day perturbative technology makes computing  $\alpha_s$-corrections
to the Wilson coefficients of nonperturbative operators feasible.
It is also understood how to treat higher-order power
corrections in a way which renders them suppressed \cite{imprec}.

High accuracy can be achieved in a comprehensive approach
where many observables are measured in $B$ decays to extract necessary
`theoretical' input parameters. It is crucial that here one can do
without relying on charm mass expansion at all, i.e.\ do not assume
charm quark to be really heavy in strong interaction mass scale. 
For reliability of the $1/m_c$ expansion is questionable. 
Already in the $1/m_Q^2$ terms one has $\frac{1}{m_c^2}
\!>\!14\frac{1}{m_b^2}$; even for the worst mass scale in the 
width expansion,   
$\frac{1}{(m_b\!-\!m_c)^2}$ is at least $8$ times smaller 
than $\frac{1}{m_c^2}$. 
There are indications \cite{chrom} that the 
nonlocal correlators
affecting meson masses can be particularly large -- a
pattern independently observed in the 't~Hooft model \cite{lebur}. This
expectation is supported by the pilot lattice study \cite{kronsim2}
which -- if taken at face value -- 
suggests a very large value of a particular combination 
$\rho_{\pi\pi}^{3\!}\!\!+\!\rho_{S}^{3}$ entering in the 
conventional approach. On the other hand, 
non-local correlators are not measured in inclusive $B$ decays.

The approach which is totally free from relying on charm mass
expansion \cite{amst} was put forward at the previous CKM-2002 Workshop at
CERN. It allows to utilize the full 
power of the comprehensive studies, and  makes use of 
a few key facts \cite{optical,motion}:\\
~\hspace*{.2em}$\bullet$ Total width to order $1/m_b^3$ is affected by
a single new Darwin operator; 
the moments also weakly depend on $\rho_{LS}^3$.\\
~\hspace*{.2em}$\bullet$ No nonlocal correlators ever enter {\it per se}.\\
~\hspace*{.2em}$\bullet$ Deviations from 
the HQ limit in the expectation values are
driven by the maximal mass scale, $2m_b$ (and are additionally
suppressed by proximity to the BPS limit); they are negligible in
practice. \\
~\hspace*{.2em}$\bullet$ Exact sum rules and inequalities which 
hold for properly defined Wilsonian parameters.

The original motivation for the precision control of strong
interaction effects in $B$ decays was accurate determination of
$|V_{cb}|$ and $|V_{ub}|$ as means to cross check the Standard
Model. For theorists it is clear, however that interesting physics
lies not only in the CKM matrix; knowledge of heavy quark masses and
nonperturbative parameters of QCD is of high importance as well.

Some of the HQ parameters like $\mu_G^2$ are known beforehand. Proper 
field-theoretic definition allows its accurate determination 
from the $B^*\!-\!B$ mass splitting:
$\mu_G^2(1\GeV)\!=\!0.35^{+.03}_{-.02}\GeV^2$ \cite{chrom}. 
A priori less certain is $\mu_\pi^2$. 
However, the inequality $\mu_\pi^2\!>\!\mu_G^2$ valid 
for any definition of
kinetic and chromomagnetic operators respecting the QCD commutation
relation $[D_j,D_k]\!=\!-ig_sG_{jk}$, 
and the corresponding sum rules 
essentially limit its range: $\mu_\pi^2(1\GeV)\!=\!0.45\!\pm\!
0.1\GeV^2$.
\vspace*{2mm}

\noindent
{\bf Lepton and hadron moments.}
Moments of the charged lepton energy in the semileptonic $B$ decays
are traditional observables to measure heavy quark parameters. 
Moments of the invariant
hadronic mass squared $M_X^2$ in semileptonic decays is 
another useful set of observables. Their utility 
follows from the observation \cite{optical} that, at least if
charm were heavy enough  
the first, second and third moments would more or less
directly yield $\La$, $\mu_\pi^2$ and 
$\rho_D^3$. Precision measurements of the $B\!\to\! X_s+\gamma$ decays
also may yield a number of constraints.
Details of the present situation 
can be found in the 
experimental review talks \cite{cassel,luth,calvi,battag}.

Let me briefly illustrate how this strategy works number-wise.
Lepton energy moments, for instance, are given 
by the following approximate expressions ($b\!\to\! u$ decays are neglected):
\bea
\nonumber
&&\mbox{\hspace*{-8mm}~}\aver{E_\ell}\!=\!1.38\GeV \!+\! 
0.38[(m_{b\!}\!-\!4.6\GeV) 
\!-\!0.7(m_{c\!}\!-\!1.15\GeV)]\\
\nonumber
&&\mbox{\hspace*{.55mm}~}
+0.03(\mu_\pi^2\!-\!0.4\GeV^2) 
-0.09(\tilde\rho_D^3\!-\!0.12\GeV^3)\;,
\rule[-2mm]{0mm}{4mm}\\
\nonumber
&&\mbox{\hspace*{-8mm}~}\aver{(E_\ell\!-\!\aver{E_\ell})^2}\!=\!0.18\GeV^2 \!+ 
0.1[(m_b\!-\!4.6\GeV) -\\
\nonumber 
&&\mbox{\hspace*{10mm}~} 0.6(m_c\!-\!1.15\GeV)] +
0.045(\mu_\pi^2\!-\!0.4\GeV^2)\\  
\nonumber 
&&\mbox{\hspace*{35mm}~}-0.06(\tilde\rho_D^3\!-\!0.12\GeV^3)\;,\\
\nonumber 
&&\mbox{\hspace*{-8mm}~}\aver{(E_\ell\!-\!\aver{E_\ell})^3} = -0.033\GeV^3 - 
0.03\,[(m_b\!-\!4.6\GeV)\\
\nonumber 
&&\mbox{\hspace*{3mm}~}
-0.8(m_c\!-\!1.15\GeV)]  
+ 0.024(\mu_\pi^2\!-\!0.4\GeV^2)
\\   
&&\mbox{\hspace*{29.5mm}~}
-0.035(\tilde\rho_D^3\!-\!0.12\GeV^3)
\label{30}
\eea
(all dimensionful factors are given in the
corresponding powers of $\GeV$). 
The moments depend basically on one and the same
combination of masses $m_b\!-\!0.65m_c$; dependence on $\mu_\pi^2$ is
rather weak. To even larger extent this applies to the CLEO's cut
moments $R_1$, $R_2$ and the ratio $R_0$ -- they depend practically on
a single combination $m_b\!-\!0.63m_c\!+\!0.3\mu_\pi^2$. The effect of
the spin-orbital average $\rho_{LS}^3$ is negligible. The uniform
dependence can be understood noting that the spectrum is mainly
determined by the parton expression, with nonperturbative effects
playing a relatively insignificant role.

In spite of this `inefficiency' of higher moments in constraining heavy 
quark parameters,
lepton moments already allow a decent determination of 
$|V_{cb}|$. Indeed, its value extracted from $\Gamma_{\rm sl}(B)$ 
has the following dependence: 
{\small \vspace*{-1mm}
\bea
\nonumber
&&\mbox{\hspace*{-8mm}~}\mbox{{\large
$\frac{|V_{cb}|}{0.042}$}} 
= 1 -0.65[(m_b\!-\!4.6\GeV)
\!-\!0.61(m_c\!-\! 1.15\GeV)] \\
\nonumber
&&\mbox{\hspace*{4.1mm}~}+0.013\,(\mu_\pi^2\!-\!0.4\GeV^2)
+\,0.1(\tilde\rho_D^3\!-\!
0.12\GeV^3)\\
\nonumber
&&\mbox{\hspace*{4.1mm}~} +0.06(\mu_G^2\!-\!0.35\GeV^2)
-0.01(\rho_{LS}^3\!+\!0.15\GeV^3) = \rule[-2mm]{0mm}{4mm}\\ 
\nonumber
&&\mbox{\hspace*{3.5mm}~}\raisebox{-.2mm}{\mbox{{\normalsize$1$}}}\:-\:
\raisebox{-.2mm}{\mbox{{\normalsize$\frac{0.65}{0.38}$}}}\,
[\aver{E_\ell}\!-\!1.38\GeV] \,-\, 0.06\, (m_c\!-\!1.15\GeV) 
\\
\nonumber
&&\mbox{\hspace*{4.1mm}~}- 0.07(\mu_\pi^2\!-\!0.4\GeV^2)
-0.05(\tilde\rho_D^3\!-\!0.12\GeV^3)\,-
\\
&&\mbox{\hspace*{5.5mm}~} 0.08(\mu_G^2\!-\!0.35\GeV^2)-
0.005\,(\rho_{LS}^3\!+\!0.15\GeV^3);
\label{34}
\eea}
\hspace*{-.4em}a combination of the parameters has been replaced by 
the first lepton moment in Eq.~(\ref{30}), and the residual 
sensitivity to $\mu_G^2$ and $\rho_{LS}^3$ is illustrated. 
The combination determining the semileptonic width is also very close!
We see that the precise value of charm mass is
irrelevant, but reasonable accuracy in $\mu_\pi^2$
and $\tilde\rho_D^3$ is required.

The first {\tt hadronic} moment takes the form 
\bea
\nonumber
&&\mbox{\hspace*{-8mm}~}
\aver{M_X^2} = 4.54\GeV^2 - 5.0\,[(m_b\!-\!4.6\GeV)-\\
\nonumber
&&\mbox{\hspace*{36mm}~}
0.62\,(m_c\!-\!1.15\GeV)]\\
&&\mbox{\hspace*{6.0mm}~}
-0.66\,(\mu_\pi^2\!-\!0.4\GeV^2)+
(\tilde\rho_D^3\!-\!0.12\GeV^3)\,,
\label{40}
\eea
i.e., again given by nearly the same combination
$m_b\!-\!0.7m_{c\!}+\!0.1\mu_\pi^2\!-\!0.2\rho_D^3$ as the lepton moment. 
Not very constraining, this provides, however a highly nontrivial check
of the HQ expansion. For example, taking the DELPHI's central value for
$\aver{M_X^2}$ we would predict $\aver{E_\ell}=1.377\GeV$, while
experimentally they obtain $\aver{E_\ell}=(1.383\pm 0.015)\GeV$ \cite{calvi}.
In this respect such a comparison is more critical than among the 
lepton moments themselves. In particular, these two first 
moments together verify the heavy
quark sum rule for $M_B\!-\!m_b$ with the accuracy about $40\MeV$!

The dependence on heavy quark parameters expectedly changes for 
higher hadronic moments:
\bea
\nonumber
&&\mbox{\hspace*{-8mm}~}
\aver{(M_{X\!}^2\!-\!\aver{M_X^2})^2}\!=\!1.2\GeV^{4\!}  
\!-\!0.003(m_b\!-\!4.6\GeV)\\
\nonumber
&& \mbox{\hspace*{6mm}~}-0.68\,(m_c\!-\!1.15\GeV) +
4.5\,(\mu_\pi^2\!-\!0.4\GeV^2)\\
\nonumber
&&\mbox{\hspace*{36mm}~} 
-5.5\,(\tilde\rho_D^3\!-\!0.12\GeV^3)\;,\\
\nonumber
&&\mbox{\hspace*{-8mm}~} 
\aver{(M_X^2\!-\!\aver{M_X^2})^3}= 4\GeV^6 +(m_b\!-\!4.6\GeV)
\\
\nonumber
&&\mbox{\hspace*{14mm}~}
-3\,(m_c\!-\!1.15\GeV) + 5\,(\mu_\pi^2\!-\!0.4\GeV^2)
\\
&&\mbox{\hspace*{34mm}~}
+13\,(\tilde\rho_D^3\!-\!0.12\GeV^3)\;.
\label{44}
\eea
Ideally, they would measure the kinetic and Darwin expectation values
separately. At present, however, we have only an approximate
evaluation and informative upper bound on $\tilde\rho_D^3$. The
current sensitivity to $\mu_\pi^2$ and $\tilde\rho_D^3$ is about
$0.1\GeV^2$ and $0.1\GeV^3$, respectively.

The experimental constraint on the combination driving $\Gamma_{\rm
sl}(B)$ in the reported DELPHI measurements appears stronger for 
the hadronic moment. Using it instead of  $\aver{E_\ell}$ we would
arrive at 
{\small
\bea
\nonumber
&&\mbox{\hspace*{-8mm}~}\mbox{{\large
$\frac{|V_{cb}|}{0.042}$}} 
\!=\! \mbox{\hspace*{-.08mm}~}\raisebox{-.2mm}{\mbox{{\normalsize$1$}}}\,+\,
0.14\,
[\aver{M_X^2}\!-\!4.54\GeV^2] \,-\, 0.03\, (m_c\!-\!1.15\GeV) \\
&& \msp{5.2}+\, 0.1\,(\mu_\pi^2\!-\!0.4\GeV^2)
 +0.1\,(\tilde\rho_D^3\!-\!0.12\GeV^3)\,.
\label{45}
\eea}
We see that measuring the second and third hadronic moments is 
an essential step in implementing the comprehensive program of extracting
$|V_{cb}|$ (see figures in M.~Calvi's talk, \cite{calvi}, and in
\cite{DELPHI}). 
Neglecting possible theoretical uncertainties in
the above relations, we get, for example, 
\beq
|V_{cb}|=0.0421\,\left(1\pm 0.01_{\rm SL\,width}  
\pm 0.015_{\rm HQ\,par}\right)
\label{46}
\eeq
from only DELPHI hadronic moments. Incorporating into the fit the full
set of moments they arrived at even smaller error interval. 

Clearly, more work -- both theoretical and experimental --
is required to fully use the potential of inclusive semileptonic decays.  
It is crucial that
this extraction carries no hidden assumptions, and at no point we
rely on $1/m_c$ expansion. Charm quark could be either heavy, or
light as strange or up quark, without deteriorating -- and rather
improving -- the accuracy.

A similar analysis can be applied to the moments with a cut on lepton
energy; in particular, CLEO has measured a few lepton energy moments for
$E_\ell \!>\! 1.5\GeV$ with unprecedented accuracy. As mentioned above, they
also fix more or less the same combination of masses and
nonperturbative parameters. The value comes out close, but does not
literally coincide with that obtained by DELPHI. This clearly
deserves further scrutiny. However, since the accuracy of theoretical
expressions is limited, in particular for relatively high cut
employed, at the moment I do not see reasons to worry. As we heard at
this Workshop \cite{cassel}, CLEO was able to do similar measurements
lowering the cut down to about $1\GeV$. We are looking forward to the
new data -- they can be treated more reliably by theory and are
expected to provide critical checks. 
\vspace*{1mm}

Due to space and time limitations I have to omit discussion of an 
important question of convergence of power expansion we employ for the
analysis of inclusive decays, and closely related to it theoretically
problem of possible violations of quark-hadron duality. Referring the
reader to the dedicated publications \cite{vadem,imprec} (an
interesting complementary discussion can also be found in the recent
CKM proceedings \cite{ckmrep}, Chapter~3, Sect.~2.3), I only
mention that the answer radically depends on what concretely is
studied, and how. Using the proper approach the effects are under good
control for total semileptonic widths, but not necessary so 
for higher hadronic
moments where some improvements may be required to achieve desirable
precision. 
\vspace*{1mm}

\noindent
{\bf Experimental cuts and hardness.}
There is a problem, however, which should not be underestimated. 
The intrinsic `{\tt hardness}' of the moments deteriorates when the cut
on $E_\ell$ is imposed. As a result, say the extraordinary
experimental 
accuracy of CLEO's $R_0$--$R_2$ cannot be even nearly
utilized by theory, whether or not the expressions we use make
this explicit.\footnote{An instructive example of how naive analysis
can miss such effects was given in \cite{uses}, Sect.~5.}

For total widths the effective energy scale parameter is generally
${\cal Q}\!=\!m_b\!-\!m_c$. 
Where OPE applies we can go beyond purely qualitative speculations about
hardness. Then 
it is typically given by ${\cal Q}\!\lsim\! \omega_{\rm
max}$, with $\omega_{\rm max}$ the threshold energy at which the
decay process kinematically disappears once $m_b$ is replaced by
$m_b\!-\!\omega$.  With the $E_\ell\!>\!E_{\rm min}$ cut then 
\beq
{\cal Q}\simeq m_b-E_{\rm min}-\sqrt{E_{\rm min}^2+m_c^2} 
\label{50}
\eeq
constituting only meager $1.25\GeV$ for 
$E_{\rm min}\!=\!1.5\GeV$, and falls
even below $1\GeV$ for the decays with $E_\ell \!>\! 1.7 \GeV$. 
A closer look reveals that such a limitation appears relevant 
for understanding the `unexpected' behavior of the first
hadronic moment with respect to the cut on $E_\ell$ reported by BaBar
last summer \cite{babarmx}. 

In  $b\!\to\! s+\gamma$ decays one has ${\cal Q}\!\simeq\!
m_b\!-\!2E_{\rm min}$, once again a rather soft scale $1.2\GeV$ if the
lower cut is placed at $E_\gamma\!=\!2\GeV$. Hence, the reliability of theory
can be questioned when one aims at maximum precision.
For higher moments the hardness further deteriorates in
either decays. A high premium 
should then be put on lowering the cuts \cite{uses}.

On the theoretical side, the higher hadronic moments can be affected
by nonperturbative physics formally scaling as powers of $1/m_b$ 
greater than $3$. At the same time, these moments are instrumental for
truly model-independent comprehensive studies of $B$
mesons; improvement is needed already for the third moment, its
expression given above is not too accurate. 
Considering alternative
kinematic variables will help to improve the convergence. 
This is related to the peculiarity of the kinematic definition of the 
invariant mass $M_X^2$:
\beq
M_X^2\!\equiv\!
(P_B\!-\!q)^2 \!=\!p_c^2+2(M_B\!-\!m_b)(m_b\!-\!q_0)
 +(M_B\!-\!m_b)^2
\label{51}
\eeq
where $p_c\equiv m_b\,v\!-\!q$ is the $c$-quark momentum as it emerges
at the quark level 
and $m_b\!-\!q_0\!\equiv\! E_c$ is its
energy; as always $q$ is the lepton pair four-momentum. In the 
standard OPE we compute separately $\Delta\!\equiv\!
p_c^2\!-\!m_c^2$ having the meaning of the final state quark virtuality,
and $E_c$. Power expansion in terms of local heavy quark operators
applies to both of them, or any their product.
Convergence of $\Delta$ or its powers is 
satisfactory; large corrections in higher moments emerge from the product
$2(M_B\!-\!m_b)E_c$. Moreover, it is powers of $\Delta$ that are directly
related to the higher-dimension expectation values. They also determine the
distribution over the off-shellness of the final-state $c$ quark, much 
in the same way as the moments of the light-cone heavy quark
distribution function shape the photon energy spectrum in $b\!\to\!
s+\gamma$.

Therefore, it is advantageous to trade the traditional
hadronic mass $M_X^2$ for the observable more closely corresponding to
the quark virtuality, defined as 
\beq
{\cal N}_X^2 \!=\!M_X^2\!-\!2\tilde\Lambda E_X\,, 
\label{52}
\eeq
where $E_X\!=\!M_{B\!}-q_0$ is the total hadronic energy in the $B$ restframe, 
and $\tilde\Lambda$ a fixed mass parameter (a constant in ${\cal
N}_X^2$ does not affect higher moments). 
Preferred values are about $\tilde\Lambda \simeq M_B\!-\!m_b(1\GeV)$ 
and can be taken $500-700\MeV$.
The higher moments 
$\aver{({\cal N}_X^2\!-\!\aver{{\cal N}_X^2})^2}$, $\aver{({\cal
N}_X^2\!-\!\aver{{\cal N}_X^2})^3}$... should enjoy better theoretical
stability. It may be possible to further improve it dividing ${\cal
N}_X^2$ by certain kinematic factors.

The kinematic variable ${\cal N}_X^2$ is not well constrained
inclusively at LEP experiments. Is it possible to measure it at
$B$-factories? This question was raised at ICHEP\,2002, 
and it seems the answer is positive; I'm grateful to 
colleagues from BaBar for
clarifying this during the $V_{xb}$ and $V_{tx}$ Workshop at SLAC. 
This possibility should be explored.
\vspace*{2mm}

\noindent
{\bf Problems for HQE?} So far I was seemingly 
quite optimistic confronting OPE-based 
heavy quark expansion for inclusive semileptonic decays to 
experiment. There are two problems, however which are often mentioned as
constituting problems for the OPE. One is a stronger than predicted 
\cite{gfit}
dependence of $\aver{M_X^2}$ on the lepton cut $E_\ell$ reported last
summer by BaBar \cite{luth}. The second is a possible deviation of
the heavy quark parameters from $b\!\to\!s+\gamma$ spectrum reportedly
deteriorating the global fit when these data are
included \cite{gfit}.

I am not sure we really witness here the problem for the theory based
on the OPE. Clearly, future confirmation of experimental results would
be necessary to draw any radical conclusion, as has been emphasized in a
number of talks. Yet, I think even the present data taken at face value
do not indicate problems with applicability of the OPE, but may rather
be attributed to certain unjustified simplifications made in the
theoretical analyses. 

The corresponding problems with the $b\!\to\!s+\gamma$ spectrum
actually had been discussed in paper \cite{uses} contributed to
the CKM\,2002 Workshop. Sect.~5 specifically pointed out the
complications arising when the cut on $E_\gamma$ is not low
enough. In this case the routinely used expressions for the moments in
the form of simple sum of perturbative and nonperturbative terms,
are not justified. This is another side of the same deteriorating 
`hardness' I discussed earlier. Theory would become
more reliable if the cut could be lowered by even $100\MeV$.

The similar problem plagues usefulness of the `global fits' like
one reported in Ref.~\cite{gfit} -- they combine accurate and well
justified theoretical expressions with those of an {\tt imprecise} nature
and treat them on parallel footing.  The literal outcome like the
overall quality of the fit may then be not too meaningful, and some
conclusions may happen to be misleading. The employed approach to treat
theoretical uncertainty was also criticized by some experimental
colleagues. While such fits are a reasonable starting point when new
data are just coming, clearly a more thoughtful and sophisticated way
to analyze the whole set of data has to be elaborated to confront data
and theory seriously and to target the ultimate precision. 

Here I would like to dwell in more detail on the first alleged
problem, the $E_\ell$-cut dependence of $\aver{M_X^2}$, since it 
attracted much
attention and caused various speculations. There may be a certain pure
experimental controversy which will eventually
lead to some change in numbers. Its concise discussion has been given by
M.~Luke \cite{luketalk}. I abstract here from such possibilities and
accept the data at face value.

On one hand, I have mentioned
that, say comparing the DELPHI's leptonic and hadronic moments one
gets an impressive agreement at the nonperturbative level. On the other
hand, the experimental measurement of $\aver{M_X^2}$ at different
$E_\ell$ performed by BaBar apparently shows a far stronger dependence
than the curve quoted as a theoretical prediction, Fig.~1. This may
look as a controversy: do we observe a triumph or failure of the OPE?

\begin{figure}
\vspace*{1.2mm}
\hbox to\hsize{\hss \hspace*{-12mm}
\includegraphics[width=0.8\hsize]{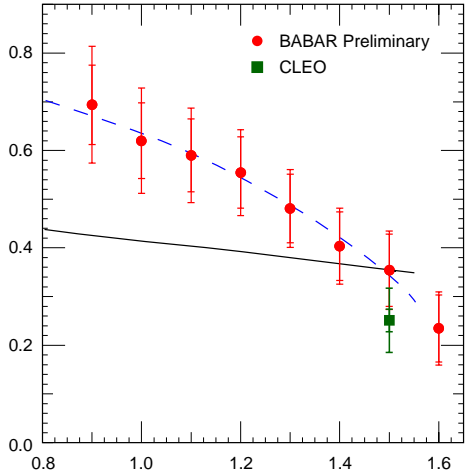}
\hss  \vspace*{-6mm}}
\caption{$\aver{M_X^2}$ vs.\ lower cut on $E_\ell$ as measured 
by BaBar \cite{babarmx}.
Nearly straight line shows their quoted theoretical prediction given
in Ref.~\cite{gfit}.}
\end{figure}

I would 
advocate the opinion that the discrepancy -- which may look 
serious -- in fact is hardly established since the
theoretical uncertainties in the evaluation of $\aver{M_X^2}$ have been
essentially underestimated here, in particular at the higher end of
$E_\ell$. (Claimed theory error bars ranged from $\pm 0.035\GeV^2$ to 
$\pm 0.05\GeV^2$ for low and high $E_\ell$, respectively \cite{gfit}.)

To give a sense of numbers without submerging into technicalities, I
recall a few relevant facts. \\
$\bullet$ OPE computes 
in first instance not the plotted
$\aver{M_X^2\!-\!M_{\bar D}^2}\!\simeq\! 0.4\GeV^2$ for which the relative
discrepancy looks of order unity, but rather
$\aver{M_X^2\!-\!(m_c\!+\!\La)^2}\!\simeq\! 1.2\GeV^2$ -- and all of
it is the
effect of strong interactions, mostly nonperturbative.\\
$\bullet$ The actual sensitivity of  $\aver{M_X^2}$ to heavy quark
parameters is illustrated by the fact that variation of $m_b$ by only
$\pm 20\MeV$ changes $\aver{M_X^2}$ by $\mp 0.1\GeV^2$, the scale of
the alleged discrepancy.\footnote{Z.\,Ligeti claimed he and
collaborators did not find such a strong dependence  in the theoretical
expressions implying that this should undermine the proposed explanation.  
I insist that this dependence is true.}\\
\hspace*{.5em}Rephrasing Ben Bradley, one may 
then suspect that the problem originates
from ``Dealing in the expressions, not necessarily in (OPE) truths''.

The sensitivity to the precise value of $m_b$ refers to varying only
$m_b$ while keeping all other heavy quark parameters we use, in
particular $m_c$ and $\mu_\pi^2$ fixed. Since, as mentioned earlier,
charm mass expansion has a questionable accuracy and this is the only
independent way to get a direct constraint on $m_b\!-\!m_c$, this is an
appropriate yardstick; even having 
$m_c$ known exactly we a priori cannot pinpoint 
the value of $m_b$ with better than $30$ to $40\MeV$.

Of course, this example helps only to illustrate the scale of
the involved theoretical uncertainties in more familiar terms. By
itself varying $m_b$ by such a small amount can shift overall theoretical 
prediction up or down, but cannot noticeably change the variation
of $\aver{M_X^2}$ with increase of the $E_\ell$ cut up to
$1.5\GeV$. Whatever $m_b$ is, it is a constant, the actual $b$ quark
mass. This might suggest the above consideration is not relevant for
reconciling theory with the data. Such a conclusion, however would
tacitly assume that other theoretical uncertainties are likewise 
independent on the placement of the cut. This would be grossly wrong. 
In fact, we know that the OPE expansion deteriorates fast when $E_\ell$
increases and must completely blow up when $E_\ell$ approaches the
borderline at a fraction of $\GeV$ below $2.15\GeV$.

The numbers following Eq.~(\ref{50}) characterizing the hardness 
qualitatively suggest the breakdown scale can be just around
$1.7\GeV$. It can be concluded in a more quantitative fashion that the
usual power expansion of $\aver{M_X^2}$ loses sense at least for cuts above
just $E_\ell\!=\!1.7\GeV$. To illustrate this, let us assume the
chromomagnetic operator vanishes, and neglect the perturbative
effects. The leading term in the OPE is then given by the kinetic
average $\mu_\pi^2$, and it contributes with the {\sf negative} sign
to $\aver{M_X^2\!-\!m_c^2}$. There is, however a positive
leading parton-like contribution $2(M_B\!-\!m_b)\aver{E_c}$. With the
canonic values of parameters, $m_b\!=\!4.6\GeV$ and 
$\mu_\pi^2\!=\!0.4\GeV^2$ 
we would obtain reasonable positive $\aver{M_X^2}\!-\!M_{\bar D}^2$ 
at low lepton cut, it would vanish for the cut near $1.7\GeV$
and, retaining only the leading terms $\aver{M_X^2}$ becomes smaller
than $M_D^2$ for higher cuts (the shift in the meson masses when 
playing with $\mu_\pi^2$ and $\mu_G^2$ should be accounted for). Clearly
this would be physically meaningless, signifying that higher
order corrections must become $100\%$ important, whatever are our
naive estimates of the next-to-leading terms.

This qualitative behavior of the OPE is quite natural. The question
then arises -- why the increasing uncertainties were not revealed in
the corresponding theoretical calculations? This is related to a
certain subtlety in applying the OPE which was missed already in the 
original estimates \cite{califold}. Those early evaluations, representing a
natural first step in the analysis, are conceptually incomplete in
this respect. The
underlying reason was addressed in Ref.~\cite{uses}, and I briefly
reiterate it here.

In $b\!\to\! c\,\ell\nu$ decays this is somewhat obscured by
three-body kinematics with massive $c$ quark. It is more transparent
for $b\!\to\! s+\gamma $ which we then can look at first. 
Let us consider a 
constrained fraction of the $B\!\to\! X_s \!+\!\gamma $ events
\beq
1\!-\!\Phi_\gamma(E) = \frac{1}{\Gamma_{bs\gamma}}\,
\int_E^{\frac{M_B}{2}} {\rm d}E_\gamma \, \frac{{\rm 
d}\Gamma_{bs\gamma}} {{\rm d}E_{\gamma}}\;.
\label{110uses}
\eeq
Ignoring the perturbative corrections, the usual $1/m_b$ expansion 
always yields the spectrum in the form of expanding around the
free-quark kinematics:
$$
\nonumber
\frac{1}{\Gamma^0_{\!bs\gamma}\!\!} \frac{{\rm d}\Gamma_{\!bs\gamma}} 
{{\rm d}E_{\gamma}} \!=\! a\,\delta(E_\gamma\!-\!\mbox{$\frac{m_b}{2}$}) +
b\,\delta'(E_\gamma\!-\!\mbox{$\frac{m_b}{2}$}) 
+ c\,
\delta''(E_\gamma\!-\!\mbox{$\frac{m_b}{2}$}) + ...
$$
where $a$, $b$, ... are given by the expectation values of local 
$b$-quark operators over $B$. Naively computing $1\!-\!\Phi_\gamma(E)$,
or spectral moments over the restricted domain 
in this  way would yield unity for any $E \!>\! \frac{m_b}{2}$ -- the
result clearly unjustified. 
The actual behavior of the rates is described by the heavy quark
distribution function. Its tail is indeed exponentially suppressed by
a typical factor $e^{-c\, {\cal Q}/\mu_{\rm hadr}}$ at ${\cal
Q}(E_\gamma)\!\gg\! \mu_{\rm hadr}$, however for ${\cal Q}(E_\gamma) 
\!\sim\! 
\mu_{\rm hadr}$  the true integral differs by terms of order $1$. 

This is an expected behavior. The point is rather that this is totally
missed in the naive OPE and in the way to gauge the theoretical
uncertainty based on it! Conceptually this is related, as illustrated
below, to the limited range of convergence of the OPE for the width,
determined in this case by the support of the heavy quark distribution
function.

The OPE for the semileptonic decays is similar in this respect. Both
the lepton spectrum itself and the moments of, say $M_X^2$ have
similar end-point singularities in the power expansion, their
resummation likewise yields a certain distribution function strongly
affecting moments at ${\cal Q} \lsim \mu_{\rm hadr}$. In 
$b\!\to\! u\,\ell\nu$ this is the domain $2E_\ell \gsim
m_b\!-\!\mu_{\rm hadr}$. 

The situation turns out more deceptive for $b\!\to\! c\,\ell\nu$. In
contrast to $b\!\to\! u$, charm mass leads to an apparent softening of
the end-point behavior where some of $\delta(E_\ell\!-\!E_{\rm max})$, 
$\delta'(E_\ell\!-\!E_{\rm max})$, ...\ are replaced by regular functions.
The latter, in fact represent the finite-width realization of the same
distributions, with the effective width $\,\sim \!m_c^2/2m_b$. If $m_c^2$
were much larger than $ \mu_{\rm hadr} m_b$ this would indeed
damp the end-point nonperturbative corrections. In practice,
however, it appears that qualitatively $m_c^2 \!\lsim\! 2\mu_{\rm hadr}
m_b$ and the effect of the nonperturbative smearing may still
dominate. Then it is not reflected in the size of the naive OPE terms
for the decay distributions. 

On the other hand, here in $b\!\to\! c\,\ell\nu$ decays these singular Wilson
coefficients do depend on the lepton cut through the `tails' of the
end-point spikes. In contrast to  $b\!\to\! s+\gamma$ where total
absence of the $E_{\rm cut}$-dependence alarms one that something is
missing, the residual dependence in theoretical expressions for 
$1/m_Q^k$ terms on $E_{\rm cut}$ masks this, and their suggested
size strongly underestimates the actual corrections in the upper part
of the spectrum.

This becomes rather obvious in the limit where $m_c^2 \!\ll\!  2\mu_{\rm hadr}
m_b$ \cite{motion}. Here the strong nonperturbative effects completely
dominate the rates or moments at $E_\ell\gsim
\frac{m_b^2-m_c^2}{2m_b}-\frac{\mu_{\rm hadr}}{2}$, while the naive
OPE terms seem to suggest the corresponding effects are still
suppressed by powers of $m_c/m_b$ there since are given by the tails
of the $\delta$-like distributions localized in the domain of
$m_c^2/2m_b$ near the end point.

This understanding reveals the proper approach to judging the
convergence, as well as the actual uncertainty of the OPE series for
the decay distributions. This follows from the consistent way of
deriving the OPE for them \cite{motion}: one should analyze the power expansion
for the whole forward transition amplitude in question, rather than
the expansion obtained for its absorptive part one routinely is
interested in.

Say, for $b\!\to\! s+\gamma$ we would need to take $T(q_0;0)$, expand
it in $1/(m_b\!-\!2q_0)$ and study convergence of the power expansion
for all $|m_b\!-\!2q_0| >m_b\!-\!2E_\gamma^{\rm max}$. For the
semileptonic width we would select \cite{vadem}
$$
\int \!{\rm d}q^2\; [q^2 \,h_1(q_0;q^2)+
\frac{1}{3}(2E_\ell(q_0\!-\!E_\ell)\!-\!\mbox{$\frac{q^2}{2}$})\,h_2(q_0;q^2)
\msp{10}
$$
\vspace*{-10mm}
\beq
\msp{35} + q^2\,(2E_\ell\!-\!q_0)\, h_3(q_0;q^2)]\; ,
\label{204}
\eeq
etc.\footnote{Convergence for hadronic $\tau$ decay widths also
should be analyzed in the similar way.} 
Computing the width directly amounts to integrating these hadronic functions
over a closed contour, which may lead to certain kinematic
cancellations of individual coefficients even if the overall result
is not convergent at all. A simple illustration is provided by the toy
function
\beq
T(q_0;0) = \frac{2}{m_b\!-\!\mu_{\rm hadr}\!-\!2q_0}
\label{206}
\eeq
for which
\beq
\frac{1}{\pi}\int_{E_{\rm min}}\!\!\! {\rm d}q_0\; \Im T(q_0;0) = 
\vartheta\left(\mbox{$\frac{m_b\!-\!\mu_{\rm hadr}}{2}$}\!-\!
E_{\rm min}\right), 
\label{208}
\eeq
while its naive power expansion in $1/m_b$ yields
$$
\frac{1}{\pi}\int_{E_{\rm min}}\!\!\! {\rm d}q_0\: \Im T(q_0;0) = 
\int_{E_{\rm min}}\!\!\! {\rm d}q_0\:
\delta(q_0\!-\!\mbox{$\frac{m_b}{2}$}) \msp{22}
$$
\beq
\msp{7}+\mbox{$\frac{\mu_{\rm hadr}}{2}$}\!
\int_{E_{\rm min}}\!\!\!\! {\rm d}q_0\;
\delta'(q_0\!-\!\mbox{$\frac{m_b}{2}$}) \,+\, ...=
\vartheta\left(\mbox{$\frac{m_b}{2}$}\!-\!E_{\rm min}\right),
\label{210}
\eeq
i.e.\ exactly $1$ for any cut below $m_b/2$. (The actual 
$b\!\to\!s+\gamma$ is not much different -- one just needs to
integrate this over $\mu_{\rm hadr}$ with the light-cone distribution
function $F(\mu_{\rm hadr})$.) The flaw in relying on 
expansion (\ref{210}) is obvious: this is not an expansion in
$\mu_{\rm hadr}/m_b$, but in powers of $\mu_{\rm hadr}/(m_b\!-\!2q_0)$
(cf.\ with energy release), and it can be used only at
$1/(m_b\!-\!2q_0)$ below the radius of convergence of the function $T(q_0;0)$
itself, which is just $m_b\!-\!2q_0 \!>\!\mu_{\rm hadr}$. The power series for
$\int \! {\rm d}q_0\:\Im T(q_0;0)$ which consists of the single,
first term, of course converge for any $q_0$, but may yield the wrong
answer.

Let me parenthetically note that the degree of convergence with
a cut in $E_\ell$ can be also
assessed by studying the OPE for moments containing additional powers of
$E_\ell\!-\!E_{\rm max}$, say $\aver{(E\!-\!E_{\rm max})^k M_X^2}$ beyond
just $\aver{M_X^2}$ we are directly interested in (here $E_{\rm max}$
is the end-point energy). However, this is a
less straightforward and somewhat more ambiguous way. 
\vspace*{1mm}

Returning to our problem, we conclude that the intrinsic 
uncertainties of the straightforward theoretical expressions were
clearly underestimated in \cite{gfit}, in particular for significant
cuts in $E_\ell$.

It is more difficult to quantify this. We know that 
$\aver{M_X^2\!-\!M_{\bar{D}}^2}$ must end up at a low value
$-0.3\GeV^2$ when one approaches the endpoint
$E_{\rm max}\!=\!\frac{M_B^2\!-\!M_D^2}{2M_B}$ since there only $D$ 
can be produced,
and $\aver{M_X^2}\!=\!M_D^2$. Since actual $\aver{M_X^2}$ is 
a smooth function of $E_{\rm cut}$, we would a priori expect a
behavior similar to the curve given by BaBar, rather than what they
showed as the theoretical prediction, at least in the right-most part of
the plot. However, at present it is difficult to evaluate the actual 
difference: the steep fall-off may start around $1.7\GeV$ or even a
little higher. Or a more gradual decrease over a wider domain can show
up for $E_\ell$ even below $1.5\GeV$, depending on the actual
behavior of higher-order power corrections. Likewise it is hard to
quantify, say the numerical difference between $\aver{M_X^2}_{E_{\rm
cut}=0} \!-\! \aver{M_X^2}_{E_{\rm cut}=1.5\,{\rm GeV}}$ and its value as given
by the first three terms in its naive $1/m_b$ expansion. It can
naturally be as large as $0.15\GeV^2$, but may turn out both smaller or
even larger. 

The {\sf practical recommendation} can be suggested as stemming from
this analysis. To extract heavy quark parameters or $|V_{cb}|$\
experiments should use as low cut on $E_\ell$ as possible in their
data sets where they are still sufficiently accurate and reliable. It
is then instructive to compare the theoretically predicted
$\aver{M_X^2}$ at higher cuts on $E_\ell$ based on these robust 
low-$E_\ell$ determinations, with direct experimental data. Since the
overall shift is very sensitive already to quark masses themselves,
and the data may be strongly correlated, the particular quantity of
interest is
\beq
\Delta M_X^2(E_{\rm cut})=\aver{M_X^2}_{E_{\rm cut}^0}\!-\!\aver{M_X^2}_{E_{\rm
cut}} \,.
\label{220}
\eeq
Theory must properly describe its slope at lower $E_{\rm cut}$, but
possibly not approaching $1.5\GeV$.
Confronting $\Delta M_X^2$ with theoretical predictions in a wider range 
we may infer a nontrivial
qualitative information about strong dynamics of heavy quarks, similar
to studying the photon spectrum in $b\!\to\! s+\gamma$.

\vspace*{2mm}

\noindent
{\bf BPS limit.}
An intriguing theoretical environment opens up if
$\mu_\pi^2(1\GeV)$ 
is eventually confirmed to be close enough to
$\mu_G^2(1\GeV)$ as currently suggested by experiment, say it does
not exceed $0.45\GeV^2$. If $\mu_\pi^2\!-\!\mu_G^2
\!\ll \!\mu_\pi^2$ it is advantageous to analyze strong dynamics 
expanding around the point
$\mu_\pi^2\!=\!\mu_G^2$ \cite{chrom}. This is not just 
one point of a continuum in
the parameter space, but a quite special `BPS' limit where the
heavy flavor  
ground state satisfies functional relations 
$\vec{\sigma}\vec{\pi}\state{B}\!=\!0$. This limit is 
remarkable in
many respects, for example, saturates the bound \cite{newsr}
$\varrho^2\!\ge\!\frac{3}{4}$ for the slope of the IW function. 
I hasten to recall that already quite some time ago there were
dedicated QCD sum rules estimates of both $\varrho^2$
\cite{bshifmanrho} and $\mu_\pi^2$ \cite{lubl} yielding literal values 
nearly at the respective lower bounds, supporting this limit.

The SV heavy quark sum rules  
place a number of important constraints on the nonperturbative
parameters. For instance, they yield a bound on the IW slope
\beq
\mu_\pi^2\!-\!\mu_G^2 \!=\!3\tilde\vep^2
(\varrho^2\!-\!\mbox{$\frac{3}{4}$}),\;\;\;\; 0.4\GeV \!\lsim\!
\tilde\vep \!\lsim\! 1\GeV\;
\label{56}
\eeq
so that $\varrho^2$ can barely reach $1$ being rather closer to
$0.85$. It is interesting that this prediction \cite{chrom} turned out
in a good agreement with the recent lattice calculation \cite{rholat}
$\varrho^2=0.83^{+.15+.24}_{-.11-.01}$. This 
leaves only a small room for the slope of the actual $B\!\to\! D^*$
formfactor, excluding values of $\hat\varrho^2$ 
in excess of $1.15\!-\!1.2$. This would be a
very constraining result for a number of experimental studies, in
particular for extrapolating the $B\to D^*$ rate to zero recoil. Since
there is a strong correlation between the extrapolated rate and the
slope, this may change the extracted value of $|V_{cb}|$. Therefore, it
is advantageous to analyze the $B\to D^* \ell\nu$ data including the
above constraint as an option, and I suggest experimental colleagues to
explore this in future analyses.

The
experimentally measured slope $\hat\rho^2$ differs from $\varrho^2$ by
heavy quark symmetry-violating corrections. The estimate by Neubert 
that  $\hat\rho^2$ is smaller than  $\varrho^2$, $\hat\rho^2\simeq
\varrho^2\!-\!0.09$ seems to be ruled out by experiment. It is not
clear if a better estimate can be made in a trustworthy way.

The whole set of the heavy quark sum rules is even more
interesting. Their constraining power depends strongly on the actual
value of $\mu_\pi^2$. When it is at the lower end of the allowed
interval, the BPS expansion appears the most effective way to analyze them. 
\vspace*{2mm}

{\bf Miracles of the BPS limit.} \hspace*{1mm}A number of useful
relation for nonperturbative parameters hold in the limit 
$\mu_\pi^2\!=\!\mu_G^2$: they include $\varrho^2\!=\!\frac{3}{4}$,
$\La\!=\!2\Si$, $\:\rho_{LS}^3\!=\!-\rho_D^3$, relations for nonlocal
corre\-lators $\rho_{\pi G}^3\!=\!-2\rho_{\pi \pi}^3$, $\rho_A^3\!+\rho_{\pi
G}^3\!=\! -(\rho_{\pi \pi}^3\!+\rho_{S}^3)$, etc.

This limit also extends a number of the heavy
flavor symmetry relations for the ground-state mesons 
to all orders in $1/m$:\\
$\bullet$ There are no formal power corrections to the relation
$M_P\!=\!m_Q+\La$ and, therefore to $m_b\!-\!m_c=M_B\!-\!M_D$.\\
$\bullet$ For the $\;B\!\to\! D\;$ amplitude the HQ limit relation between
the two formfactors 
\beq
f_-(q^2)=-\frac{M_B\!-\!M_D}{M_B+M_D}\; f_+(q^2)
\label{106}
\eeq
does not receive power corrections.\\
$\bullet$ For the zero-recoil $\;B\!\to\! D\;$ amplitude all
$\delta_{1/m^k}$ terms vanish.\\
$\bullet$ For the zero-recoil amplitude $f_+$ with massless leptons
\beq
f_+((M_B\!-\!M_D)^2)=\frac{M_B+M_D}{2\sqrt{M_B M_D}}
\label{108}
\eeq
to all orders in $1/m_Q$.\\
$\bullet$ At arbitrary velocity power corrections in $\;B\!\to\! D\;$
vanish,
\beq
f_+(q^2)=\frac{M_B+M_D}{2\sqrt{M_B M_D}} \;\, 
\xi\!\left(\mbox{$\frac{M_B^2+M_D^2-
\raisebox{.6pt}{\mbox{{\normalsize $q^2$}}}}{2M_BM_D}$}\right)
\label{110}
\eeq
so that the $\;B\!\to\! D\;$ decay rate directly yields 
Isgur-Wise function $\xi(w)$.\\
It is interesting that experimentally the slope of the  $\;B\!\to\!
D\;$ amplitude is 
indeed smaller centering around $\hat\rho_{(D)}^2\!\approx \!1.15$ 
\cite{ckmrep} 
indicating qualitative agreement with the BPS regime.

What about the $\;B\!\to\! D^*\,$ amplitude, are the corrections
suppressed as well? Unfortunately, the answer is negative. The
structure of power corrections indeed simplifies in the BPS limit,
however $\delta_{1/m^2}$,  $\delta_{1/m^3}$ are still very
significant \cite{chrom}, and the literal estimate for $F_{D^*}(0)$
falls even below $0.9$. Likewise, we expect too significant
corrections to the shape of the $B\!\to\! D^*$ formfactors. Heavy
quark {\tt spin} symmetry controlling these transitions
seems to be violently affected by strong interactions for charm.

A physical clarification must be added at this point. Absence of all
power corrections in $1/m_Q$ for certain relations may be naively
interpreted as implying that they would hold for arbitrary, even small
quark masses, say in $B\!\to\!K$ transitions. This is not correct,
though, for the statement refers only to a particular fixed order in
$1/m_Q$ expansion in the strict BPS limit. In fact the relations
become more and more accurate approaching this limit only above 
a certain  
mass scale of order $\Lam$, while below it their violation is of order
unity regardless of proximity of the heavy quark ground state to BPS.
\vspace*{2mm} 

{\bf Quantifying~deviations~from~BPS.} \hspace*{1mm}The BPS limit
cannot be exact in actual QCD, and it is important to understand the 
scale of violations of its predictions. 
The dimensionless parameter describing the deviation from BPS is
\beq
\beta =\|\pi_0^{-1}(\vec\sigma\vec\pi)\,\state{B}\| 
\equiv
\mbox{$\sqrt{3\!\left(\varrho^{2}\!-\!\frac{3}{4}\right)} =
3
\left(\sum \rule{0mm}{3mm} \right.
\mbox{\hspace*{-3.5mm}\raisebox{-2.2mm}{{\tiny $n$}}\hspace*{.35mm}}
 \left.\,|\tau_{1/2}^{(n)}|^{2}\right)^{\frac{1}{2}}$}.
\label{112}
\eeq
Numerically $\,\beta\,$ is not too small, similar in size to generic
$\,1/m_{c}\,$ expansion parameter, and the relations violated to
order $\beta$ may in practice be more of a qualitative nature only.
However, 
$\mu_\pi^2\!-\!\mu_G^2 \propto \beta^2$ can be good enough. Moreover,
we can count together powers of $1/m_c$ and 
$\beta$ to judge the real quality of a particular HQ relation.
Therefore understanding at which order in $\beta$ the BPS
relations get corrections is required. In fact, we need 
classification in powers
of $\beta$  to {\tt all} orders in $1/m_Q$.

Relations (\ref{106}) and (\ref{110}) for the $B\!\to\!D$
amplitudes at arbitrary velocity can get first order corrections in
$\beta$. Thus they may be not very accurate. The same refers to
equality of $\rho_{\pi G}^3$ and $-2\rho_{\pi\pi}^3$. The other
relations mentioned for heavy quark parameter are accurate up to 
order $\beta^2$.
The other important BPS relations hold up to order $\beta^2$ as well:\\
$\bullet$ $M_B\!-\!M_D=m_b\!-\!m_c$ and $M_D=m_c\!+\!\La$ \\
$\bullet$ Zero recoil matrix element $\matel{D}{\bar{c}\gamma_0 b}{B}$
is unity up to ${\cal O}(\beta^2)$\\
$\bullet$ Experimentally measured $B\!\to\!D$ formfactor $f_+$ near
zero recoil receives only second-order corrections in $\beta$ to all
orders in $1/m_Q$:
\beq
f_+\left((M_B\!-\!M_D)^2\right) = \frac{M_B\!+\!M_D}{2\sqrt{M_BM_D}} \;\,
+ {\cal O}(\beta^2)\;.
\label{116}
\eeq
This is an analogue of the Ademollo-Gatto theorem for the BPS
expansion. 

The similar statement then applies to $f_-$ as well, and the HQ limit
prediction for $f_-/f_+$ must be quite accurate near zero recoil. It can be
experimentally checked in the decays $B\!\to\!D\,\tau\nu_\tau$.

As a practical application of the results based on the BPS expansion, 
one can calculate the $B\!\to\!D$ decay amplitude near zero recoil to
use this channel for the model-independent extraction of $|V_{cb}|$ in
future high-luminosity experiments. For power corrections we
have 
$$
\frac{M_B\!+\!M_D}{2\sqrt{M_BM_D}}\; f_+\left((M_B\!-\!M_D)^2\right) \;=\; 1
\; + \msp{31.3}
$$ 
\vspace*{-6mm}
\beq
\msp{19} \left(
\frac{\La}{2}
\!-\!\overline\Sigma\right) 
\left(\frac{1}{m_c}\!-\!\frac{1}{m_b}\right)\mbox{{\Large $
\frac{{M_B\!-\!M_D}}{M_B+M_D}$}}-
\mbox{${\cal O}\left(
\frac{1}{m_Q^2}\!\right)$}\; .
\label{124}
\eeq
We see that this indeed is of the second order in $\beta$. Moreover, 
$\La\!-\!2\Si$ is well constrained through $\mu_\pi^2\!-\!\mu_G^2$ by
spin sum rules. Including perturbative corrections (which should be
calculated  in the proper renormalization scheme respecting BPS regime) 
we arrive at the estimate 
\beq
\frac{2\sqrt{M_B M_D}}{M_B+M_D}\; f_+(0) = 1.03\pm 0.025 \msp{10}
\label{126}
\eeq
It is valid through order $\beta^2\frac{1}{m_c}$ accounting for all
powers of $1/m_Q$ to order $\beta^1$. Assuming the counting rules
suggested above this corresponds to the precision up to $1/m_Q^4$,
essentially better than ``gold-plated'' $B\!\to\!D^*$  formfactor where
already $1/m_Q^2$ terms are large and not well known. Therefore, the
estimate (\ref{126}) must be quite accurate. In fact, the major source
of the uncertainty seems to be perturbative corrections, which can be
refined in the straightforward way compared to decade-old
calculations.

\vspace*{1mm}
\noindent
{\bf Conclusions.} Experiment has entered a new era of exploring $B$
physics at the nonperturbative level, with qualitative improvement in
$|V_{cb}|$. The comprehensive approach will allow to reach a percent
level of reliable accuracy in translating $\Gamma_{\rm sl}(B)$ to
$|V_{cb}|$. Recent experiments have set solid grounds for dedicated
future studies at $B$ factories. We have observed a nontrivial
consistency between quite different measurements, and between
experiment and QCD-based theory, at the nonperturbative level.

There are obvious lessons to infer. Experiment must strive to weaken
the cuts in inclusive measurements used in extracting $|V_{cb}|$. 
Close attention should be paid to higher moments or their special
combinations, as well as exploring complementary kinematic observables.

The theory of heavy quark decays is now a mature branch of
QCD. Recent experimental studies of inclusive decays yielded valuable
information crucial -- through the comprehensive application of all
the 
elements of the  heavy quark expansion -- for a number of exclusive 
decays as well. This signifies an important new stage of the heavy quark
theory, since only a few years ago exclusive and inclusive
decays were often viewed as largely separated, if not as antipodes 
theory-wise. 

Generally speaking, there is ample evidence that heavy quark symmetry
undergoes significant nonperturbative corrections for charm
hadrons. However, there appears a class of practically relevant
relations which remain robust. They are limited to the ground-state
pseudoscalar $B$ and $D$ mesons, but do not include spin symmetry in
the charm sector.

The accuracy of these new relations based on the proximity to the
``BPS limit'' strongly depends on the actual size of the kinetic
expectation value in $B$ mesons, $\mu_\pi^2(1\GeV)$. The experiment
must verify it with maximal possible accuracy and fidelity, without
relying on ad hoc assumptions often made in the past with limited data
available. This can be performed through the inclusive decays already
in current experiments. If its value is confirmed not to 
exceed $0.45\GeV^2$, the $B\!\to\!D$ decays can be reliably treated by
theory, and the estimate ${\cal F}_+(0)\simeq 1.03$ can provide a good
alternative element in the comprehensive program of model-independent
extraction of $|V_{cb}|$. 

There are many other important consequences of the BPS regime. The
slope of the IW function must be close to unity, and actually below
it. The related constraints on the slope $\hat\rho^2$ of the 
experimentally observed combination of $B\!\to\!D^*$ formfactors
should be incorporated in the fits aiming at extrapolating the rate to
zero recoil. Finally, I think that the semileptonic decays with $\tau$
lepton and $\nu_\tau$ have an interesting potential for both inclusive
and exclusive decays at high-luminosity machines.

\vspace*{1mm}

{\bf Acknowledgments:} It is a pleasure to thank 
M.\,Artuso, M.\,Battaglia, O.\,Buchmueller, I.\,Bigi,~M.\,Calvi,~P.\,Gam\-bino,
U.\,Langenegger, V.\,Luth, P.\,Roudeau, M.~Shifman and A.~Vainshtein
for helpful discus\-sions. 
This work was supported in part by the NSF under grant number PHY-0087419.

\end{document}